\documentclass[aps,apl,twocolumn,groupedaddress,amsmath]{revtex4}
\usepackage[latin1]{inputenc}    
\usepackage{graphicx}   

\begin{document}
 \title{Measuring current by counting electrons in a nanowire quantum dot}
 \author{S.~Gustavsson\footnote{These authors contributed equally to this work.}}
 \author{I.~Shorubalko$^*$}
 \author{R.~Leturcq}
 \author{S.~Sch\"on}
 \author{K.~Ensslin}
 \affiliation {Solid State Physics Laboratory, ETH Z\"urich, CH-8093 Z\"urich,
 Switzerland}

\date{\today}

\begin{abstract}
We measure current by counting single electrons tunneling through an
InAs nanowire quantum dot. The charge detector is realized by
fabricating a quantum point contact in close vicinity to the
nanowire. The results based on electron counting compare well to a
direct measurements of the quantum dot current, when taking the
finite bandwidth of the detector into account. The ability to detect
single electrons also opens up possibilities for manipulating and
detecting individual spins in nanowire quantum dots.
\end{abstract}

\maketitle

A highly-sensitive charge detector is a powerful tool for probing
electronic properties of mesoscopic structures.
In contrast to conventional transport measurement, the system under
investigation does not need to be connected to leads. This makes the
measurement technique low-invasive and allows charge transitions
within the nanostructure to be investigated \cite{petta:2004}.
By adding time resolution to the detector, tunneling of individual
electrons can be detected in real-time \cite{LuW:2003}.
This provides the possibility to extract statistics for the
tunneling electrons and to probe electron-electron correlations
\cite{gustavsson:2005, fujisawa:2006}, as well as for determining
electron spin dynamics \cite{elzNat:2004, petta:2005}.

Another possible application of time-resolved charge detection is to
use it as a metrology standard for current. Bylander \emph{et al}
experimentally verified the fundamental relation $I=e\,f$  by
relating a highly-correlated current $I$ through an array of tunnel
junctions to the frequency response $f$ of a single-electron
transistor \cite{bylander:2005}.
In this work, we combine a quantum dot (QD) formed in a
semiconductor nanowire with a quantum point contact (QPC) acting as
the charge detector. The large energy scales of the nanowire QD
enable operation at $T=4~\mathrm{K}$ and allow the QPC to be
operated at larger bias voltages compared to GaAs QDs
\cite{gustavssonPRL:2007}. This together with the high sensitivity
of the detector make time-resolved single-electron detection
possible in a regime where we can simultaneously measure the QD
current with a conventional current meter. In this way, we count
electrons one by one and make direct comparisons to the measured
current. We find that the current measured by counting is lower than
the one measured with conventional techniques. The difference can be
quantitatively accounted for by considering the electrons missed
because of the limited bandwidth of the charge detector, which is a
known quantity \cite{naaman:2006}.







InAs nanowires are catalytically grown by metal-organic vapor phase
epitaxy (the detailed recipe is described in \cite{chimia:2006}). An
InAs nanowire is deposited on top of a shallow (37 nm) AlGaAs/GaAs
heterostructure based two-dimensional electron gas (2DEG). The QD in
the InAs nanowire and a QPC in the underlying 2DEG are defined in a
single etching step using patterned electron beam resist as an etch
mask. This method guarantees perfect alignment as well as strong
coupling between the two devices \cite{ish_unpublished:2007}.

%

\begin{figure}[b]
\centering
 \includegraphics[width=\columnwidth]{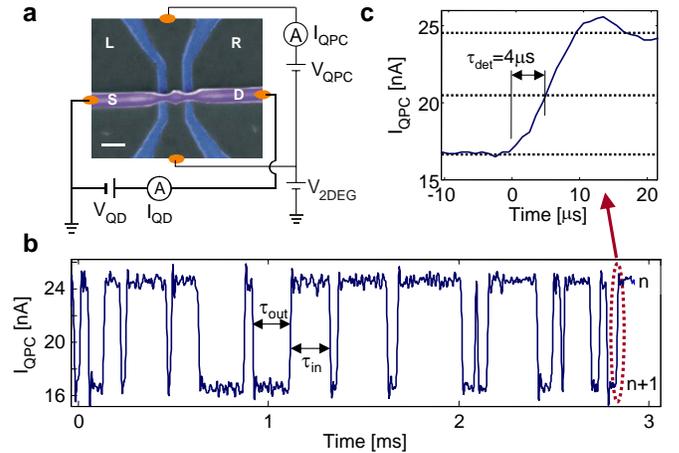}
 \caption{(color online) (a) SEM image of the device. The quantum dot is formed
 in the nanowire, with the quantum point contact located in the 2DEG
 directly beneath the QD. (b) Typical time trace of the QPC
 conductance, showing a few electrons tunneling into and out of the
 QD. The upper level corresponds to a situation with $n$ electrons
 on the QD.
 (c) Rise time of the detector, defined as the time needed for
 the current to cross the midline between current levels belonging to the $n$ and $n+1$
 electron states.
 }
\label{fig:fig1}
\end{figure}

Figure~\ref{fig:fig1}(a) shows a scanning electron microscope (SEM)
image of a device similar to the one used in the measurements. The
QD is defined by the etched constrictions in the nanowire between S
and D. The QPC is formed between the two etched trenches that
separates it from the rest of the 2DEG.
The regions marked by L and R are used as side gates to control the
QD population and to tune the coupling between the QD and the source
and drain leads. In the experiment, the QPC was biased with a DC
voltage of $V_\mathrm{QPC} = 1~\mathrm{mV}$.
In addition, a voltage was applied to the 2DEG on both sides of the
QPC to compensate for the shift in QPC potential when changing the
voltages on gates L, R. The bias of the QPC was kept smaller than
the single-level spacing of the QD to avoid QD excitations due to
photon absorbtion \cite{gustavssonPRL:2007}.
The measurements presented here were performed at a temperature of
1.7 K, but we have tested that the setup produces similar results at
$T=4~\mathrm{K}$.

The charge detector is implemented by operating the QPC at the slope
below the first plateau and continuously monitoring its conductance
\cite{field:1993}. Due to strong electrostatic coupling between the
QD and the QPC, an electron entering or leaving the QD will shift
the QPC potential and thereby change its conductance. Figure
\ref{fig:fig1}(b) shows a typical example trace of the QPC current.
Coulomb blockade prohibits the QD to hold more than one excess
electron. The two current levels in the figure corresponds to $n$
and $n+1$ electrons on the QD, respectively.
Transitions between the levels relate directly to an electron
tunneling into or out of the QD \cite{schleser:2004,
vandersypen:2004, fujisawa:2004}. The times $\tau_\mathrm{in}$,
$\tau_\mathrm{out}$ describe the times needed to tunnel into and out
of the QD.
The change in QPC conductance when adding an electron to the QD is
around $30~\%$. This is large change compared to most top-gate
defined GaAs QDs, where the relative QPC conductance change is
typically around one percent \cite{vandersypen:2004, reilly:2007}.

The comparatively large signal allows us to increase the bandwidth
of the detector. Figure~\ref{fig:fig1}(c) shows the rise time of the
detector signal. Setting the threshold for event detection in the
middle between the two levels, we find that the detector has a time
resolution of $\tau_\mathrm{det}=4~\mathrm{\mu s}$. The time scale
corresponds to a maximal detectable current of $\sim\! e
/(4\,\tau_\mathrm{det}) \sim 10~\mathrm{fA}$; tunneling occuring on
a faster timescale can not be resolved by the detector.
In the present setup, the bandwidth is limited by the low-pass
filter due to the capacitance of the cables between the sample and
the room-temperature amplifier. The bandwidth can be greatly
enhanced by using a cold amplifier \cite{vink:2007} or an rf-QPC
setup \cite{muller:2007, reilly:2007, cassidy:2007}.

\begin{figure}[tb]
\centering
 \includegraphics[width=\columnwidth]{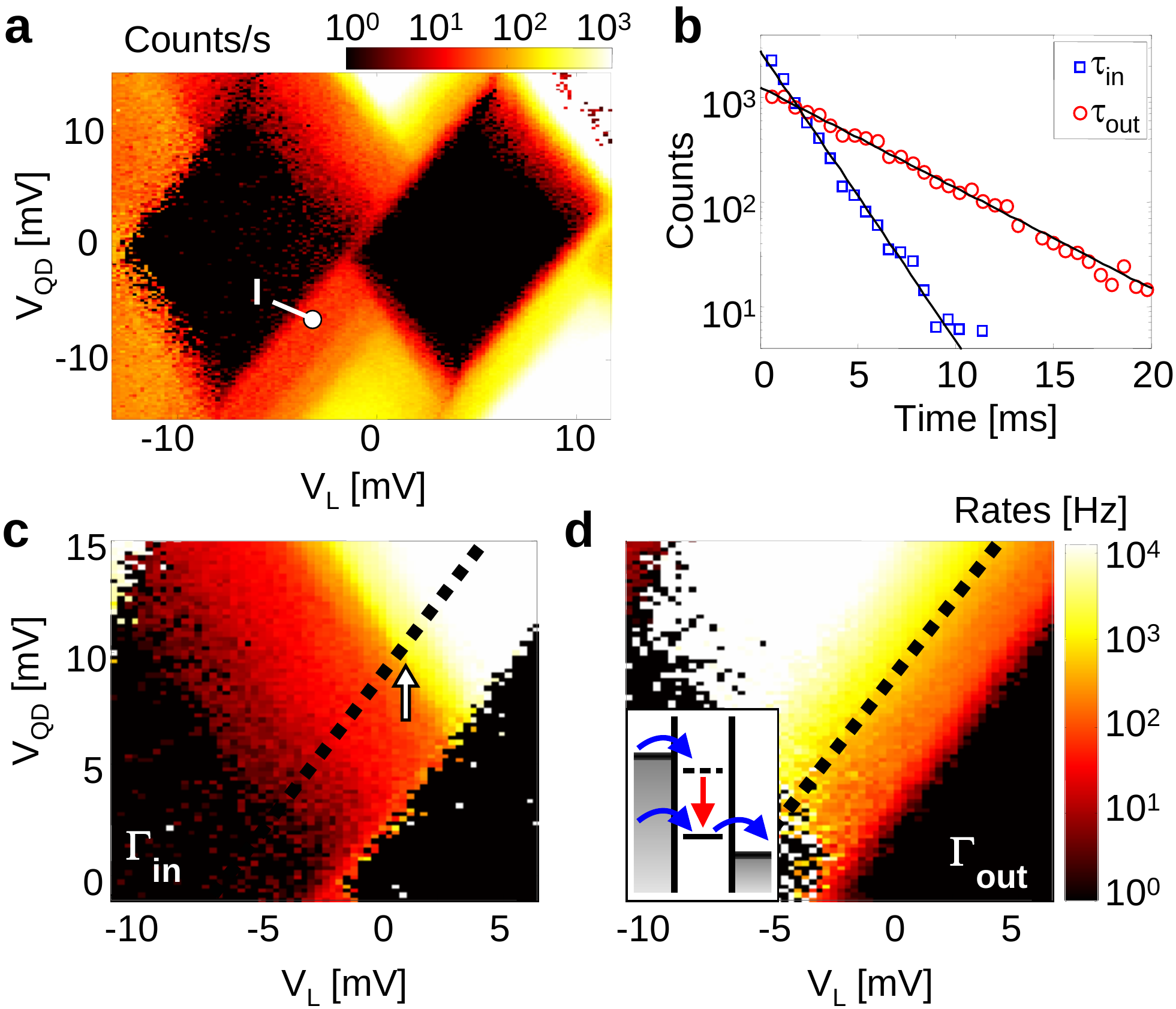}
 \caption{(color online) (a) Coulomb diamonds measured by counting electrons
 entering and leaving the QD. To compensate for changes in the QPC potential, we
 set $V_\mathrm{R}=V_\mathrm{L}-50~\mathrm{mV}$ and $V_\mathrm{2DEG}=0.7\,V_\mathrm{L}+177~\mathrm{mV}$.
 (b) Distribution of tunneling times,
 taken at the position marked by I in (a). The solid lines are fits
 to Eq. (\ref{eq:expDecay}) in the text, with fitting parameters
 $\Gamma_\mathrm{in}=640~\mathrm{Hz}$ and
 $\Gamma_\mathrm{out}=220~\mathrm{Hz}$.
 (c, d) $\Gamma_\mathrm{in}$, $\Gamma_\mathrm{out}$ for the upper-middle
 part of (a). The arrow in (c) mark the position where an excited state enters the transport
 window. The inset of (d) depicts the energy levels of the system at
 the position marked by the arrow in (c).
 }
\label{fig:fig2}
\end{figure}

To characterize the system, we first tune the tunneling rates to be
much slower than the time resolution of the detector.
Figure~\ref{fig:fig2}(b) shows Coulomb diamonds measurements for the
QD, measured by counting electrons from traces such as the one shown
in Fig.~\ref{fig:fig1}(b). The large charging energy
($E_C=12~\mathrm{meV}$) is due to the small size of the QD.
In the regime of single-level transport, the tunneling times
$\tau_\mathrm{in/out}$ are expected to follow an exponential
distribution
\begin{equation}\label{eq:expDecay}
p_{\mathrm{in/out}}(t) \mathrm{dt} = \Gamma_{\mathrm{in/out}}
\mathrm{e}^{-\Gamma_{\mathrm{in/out}} t} \mathrm{dt}.
\end{equation}
Figure~\ref{fig:fig2}(b) shows the measured distribution of the
tunneling times, taken at the point marked by I in
Fig.~\ref{fig:fig2}(a). The solid lines are fits to
Eq.~(\ref{eq:expDecay}), with $\Gamma_\mathrm{in} = 640~\mathrm{Hz}$
and $\Gamma_\mathrm{out} = 220~\mathrm{Hz}$.

\begin{figure}[tb]
\centering
 \includegraphics[width=\columnwidth]{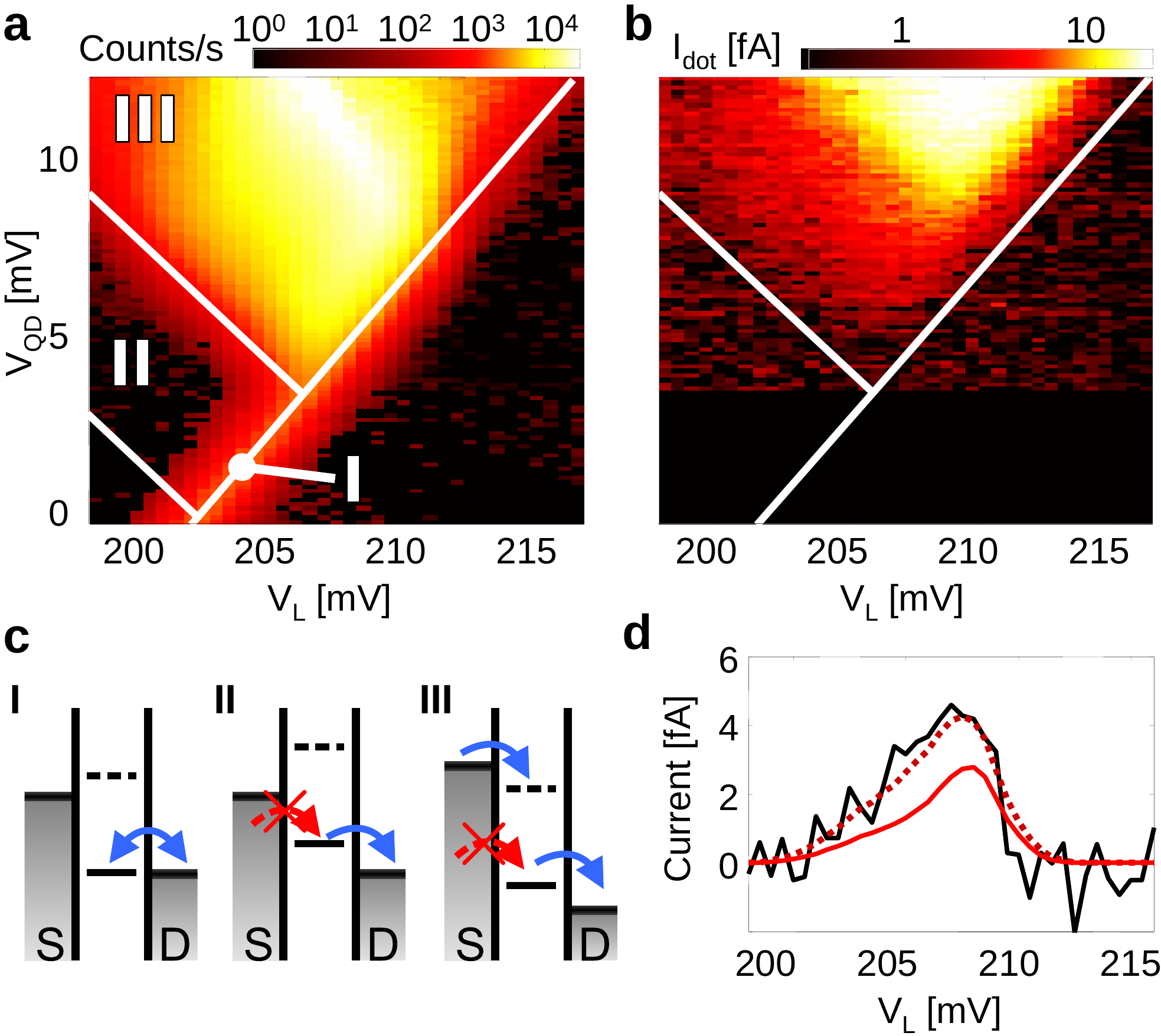}
 \caption{(color online) (a) Electron count rate, measured with the QD in a more
 open regime. (b) QD current for the same region as in (a),
 measurement with a conventional current meter. (c) Energy level
 diagrams for the three regions marked in (a). In I, the count rate
 is due to equilibrium fluctuations between the QD and the drain
 lead. In region II, transport is blocked due to weak coupling
 between the QD ground state and the source lead. In III, a more
 strongly coupled excited state is available for transport and the
 current through QD is strongly increased.
 (d) Cross section of the colormaps in (a,b), taken at $V_{QD}=7.1~\mathrm{mV}$.
 The black curve is the current measured with current meter, while
 the gray is measured by counting electrons. The dashed line is the
 counting signal when compensating for the limited bandwidth of the detector.
 }
\label{fig:fig3}
\end{figure}

In Figs.~\ref{fig:fig2}(c, d), we plot the separate tunneling rates
$\Gamma_\mathrm{in/out}$ for the upper part of the middle diamond in
Fig.~\ref{fig:fig2}(a). Going upwards along the dashed line in
Figs.~\ref{fig:fig2}(c, d), the Fermi level of the source lead is
raised while the potential of the drain and the QD is kept constant.
At $V_L=-1~\mathrm{mV}$ (marked by an arrow in the figure), there is
a distinct step in $\Gamma_\mathrm{in}$ as the source lead is raised
above an excited state of the QD. At the same time, the rate for
tunneling out measured along the same line stays constant. We
attribute this to fast relaxation of the excited state, so that the
tunneling out process always occurs through the QD ground state
\cite{gustavsson:2006}. The situation is depicted in the inset of
Fig.~\ref{fig:fig2}(d).
The results of Fig.~\ref{fig:fig2} demonstrate the stability and
high level of control in the system and prove that the electron
tunneling detected by the QPC originate from a QD formed in the
nanowire.


In the following, we present measurements in a regime where the
barriers between the QD and the leads are opened up to allow the QD
current to be measured with a conventional current meter.
Figure~\ref{fig:fig3}(a) shows the count rate for the positive bias
part of a Coulomb diamond.
In this regime, the ground state of the QD is weakly coupled to the
source lead. The measurement shows equilibrium fluctuations between
the QD and the drain lead [region I in Fig.~3(a, c)], but almost no
counts inside the region marked by II. As the bias is further
increased, the first excited state is available for transport and
the count rate is increased (region III).
Figure~\ref{fig:fig3}(c) displays the current through the QD for the
same region as in (a), measured with a conventional
current-to-voltage (I-V) converter. We only see current inside the
regime corresponding to region III in Fig.~\ref{fig:fig3}(a). This
is expected since the current measurement in contrast to the charge
detector is directional; charge fluctuations between the QD and the
drain lead as depicted in region I of Figs.~\ref{fig:fig3}(a, c)
will not contribute to a net current flow. The discrepancies between
the counting and current signal in the upper-right corner of
Figs.~\ref{fig:fig3}(a, c) are due to the limited time resolution of
the detector; a second excited state entering the bias window makes
the tunneling-in rate exceed the detector bandwidth.

In Fig.~\ref{fig:fig3}(d), we plot the QD current together with the
electron count rate for fixed bias on the QD ($V_\mathrm{QD} =
7.1~\mathrm{mV}$). The count rate has been converted to current
using $I = e/\langle \tau_\mathrm{in}+\tau_\mathrm{out}\rangle$.
Even though the two curves show qualitatively the same behavior, the
charge detector registers a current which is $\sim\!30\%$ lower than
the one measured by the I-V converter.
We attribute the difference to the limited bandwidth of the charge
detector. Tunneling events occurring on a timescale on the order of
or faster than the time resolution of the detector are less likely
to be detected, which modifies the measured statistics
\cite{naaman:2006, gustavsson:2007}. However, knowing the detection
time and assuming that Eq.~(\ref{eq:expDecay}) correctly describes
the distribution of tunneling times, we can estimate the number of
electrons missed by the detector. Following the ideas of Naaman and
Aumentado \cite{naaman:2006}, we find that the current is given by
\begin{equation}\label{eq:currFiniteBW}
 I = e / \left(\left(\tau^*_\mathrm{in}+\tau^*_\mathrm{out}\right)
 \left( 1-\tau_\mathrm{det}
 \frac{\tau^*_\mathrm{in}+\tau^*_\mathrm{out}}
 {\tau^*_\mathrm{in} \, \tau^*_\mathrm{out}}
 \right)
 \right).
\end{equation}
Here, $\tau^*_\mathrm{in}$, $\tau^*_\mathrm{out}$ are average
tunneling times extracted from the measurement. The results of
Eq.~(\ref{eq:currFiniteBW}) is shown as the dashed line in
Fig.~\ref{fig:fig3}(d), with $\tau_\mathrm{det}=4~\mathrm{\mu s}$ as
extracted from Fig.~\ref{fig:fig1}(c).
The current calculated taking the finite bandwidth into account
agrees very well with current measured with the I-V converter. We
emphasize that the curve does not include any free parameters, since
the detection time is determined separately using the method shown
in Fig.~\ref{fig:fig1}(c).
It should be noted that the current measured by counting is
determined with much higher precision than with conventional
methods. The signal of the I-V converter was integrated for
$10~\mathrm{s}$ at each point, yielding a resolution of
$\sim\!1~\mathrm{fA}$. Also, the signal had to be carefully
compensated for amplifier drift. On the other hand, the counting
signal was measured for $0.2~\mathrm{s}$,
giving a standard deviation of only $70~\mathrm{aA}$. 

We have demonstrated current measurements by counting electrons in a
nanowire quantum dot.
The insensitiveness to drift and the high precision of the counting
procedure demonstrate big advantages of electron counting compared
to conventional current measurement techniques.
The measurements were performed at a temperature of 1.7 K, but the
large charging energy and single-level spacing of the quantum dot
allows operation even at $T=4~\mathrm{K}$.
By incorporating the sample into a radio-frequency setup, we
estimate that the detection bandwidth can be increased by at least
three orders of magnitude \cite{muller:2007, reilly:2007,
cassidy:2007}.
Combining charge readout with fast gate-pulsing techniques opens up
the possibility of investigating and manipulating individual spins
in nanowire quantum dots \cite{elzNat:2004, petta:2005}.

\bibliographystyle{apsrev}
\bibliography{CountingNanowire}

\end{document}